\documentclass[useAMS, usenatbib]{mn2e} 
\usepackage{graphicx}
\usepackage{natbib}
\newcommand{\hii}{H\,{\sevensize II} }
\newcommand{\msun}{\mathrm{M_\odot} }
\title[Periodic masers in G9.62+0.20E]{Periodic class II methanol masers
  in G9.62+0.20E} \author[DJ van der Walt et al.]{DJ van der
  Walt$^{1}$\thanks{E-mail: johan.vanderwalt@nwu.ac.za}, S
  Goedhart$^{2}$, MJ Gaylard$^{2}$ \\ $^{1}$ Unit for Space Physics,
  North-West University, Private Bag X6001, Potchefstroom, South Africa \\ $^{2}$
  Hartebeesthoek Radio Astronomy Observatory, PO Box 443, Krugersdorp,
  South Africa}

\begin{document}

\date{Accepted. Received; in original
form }

\pagerange{\pageref{firstpage}--\pageref{lastpage}} \pubyear{2002}

\maketitle

\label{firstpage}

\begin{abstract}
We present the light curves of the 6.7 and 12.2 GHz methanol masers in
the star forming region G9.62+0.20E for a time span of more than 2600
days. The earlier reported period of 244 days is confirmed. The results
of monitoring the 107 GHz methanol maser for two flares are also
presented. The results show that flaring occurs in all three masing
transitions. It is shown that the average flare profiles of the three
masing transitions are similar. The 12.2 GHz masers are the most
variable of the three masers with the largest relative amplitude having
a value of 2.4. The flux densities for the different masing transitions
are found to return to the same level during the low phase of the
masers, suggesting that the source of the periodic flaring is situated
outside the masing region, and that the physical conditions in the
masing region are relatively stable. On the basis of the shape of the
light curve we excluded stellar pulsations as the underlying mechanism
for the periodicity. It is argued that a colliding wind binary can
account for the observed periodicity and provide a mechanism to
qualitatively explain periodicity in the seed photon flux and/or the
pumping radiation field. It is also argued that the dust cooling time is
too short to explain the decay time of about 100 days of the maser
flare. A further analysis has shown that for the intervals from days 48
to 66 and from days 67 to 135 the decay of the maser light curve can be
interpreted as due to the recombination of a thermal hydrogen plasma
with densities of approximately $1.6 \times 10^6~\mathrm{cm^{-3}}$ and
$6.0 \times 10^5~\mathrm{cm^{-3}}$ respectively.
 
\end{abstract}

\begin{keywords}
ISM:clouds -- ISM:H{\sc ii} regions -- ISM:molecules
\end{keywords}

\section{Introduction}
\label{sec-intro}

Although class II methanol masers are now generally accepted to be
exclusively associated with massive star forming regions \citep[see
  eg.][]{ellingsen2006, xu2008}, it is not yet clear how much can be
learned about the star formation environment from the masers. Underlying
this uncertainty is the fact that it is difficult to determine where in
the circumstellar environment the masers operate based solely on the
spatial distribution and velocity structure of the masers
\citep{beuther2002}. In the past very little attention has been given to
the time domain aspect of the masers. Various authors \citep[see
  eg.][]{macleod1993, caswell1995, moscadelli1996, goedhart2002,
  niezurawska2002, goedhart2003, goedhart2004, goedhart2005a} have
studied the variability of the 6.7 and 12.2 GHz masers over various
periods of time and have found that variability is a common feature
amongst these masers. The most systematic study on the variability of
6.7 GHz masers to date was done by \citet{goedhart2004} using the
Hartebeesthoek Radio Astronomy Observatory (HartRAO) 26-m telescope,
revealing a wide variety of time-dependent behaviour in the masers. Six
out of their list of 54 sources were identified as periodic, with
periods ranging from 133 to 504 days \citep{goedhart2007}. Of particular
interest here are the periodic masers in the star forming region
G9.62+0.20E which show repeated flaring activity with a period of about
244 days \citep{goedhart2003, gaylard2007}

The star forming region G9.62+0.20 has been the object of study by a
number of authors. \citet{garay1993} mapped this region with the VLA at
1.5, 4.9 and 15.0 GHz and identified five extended, compact and
ultracompact (UC) \hii regions, labelled A - E. A striking feature in
this star forming complex is the near perfect alignment of various
tracers (masers, hot molecular clumps and UC \hii regions) of star
forming activity extending between components C and D, with component E
lying between C and D \citep[see][]{hofner1994}. \citet{hofner1994}
interpreted the overall morphology of this star forming region as
suggestive of induced star formation progressing from the most evolved
\hii region (component A) with the most recently formed stars located in
the linear structure.  The methanol masers in the star forming
  complex are associated with components D and E and are in the
  literature also referred to as G9.619+0.193 and G9.621+0.196
  respectively \citep{phillips1998}. For component E, the masers have a
  maximum linear extent of about 566 AU (2.75 milliparsec) based on the
  12.2 GHz observations of \citet{goedhart2005b} and using a distance of
  5.15 kpc (A. Sanna et al., in preparation).

Component E shows all the signs of a very early phase of massive star
formation. According to \citet{hofner1996} the continuum emission
follows a power law with a spectral index of $1.1 \pm 0.3$ between 2 cm
and 2.7 mm, which can be interpreted to be due to either an ionized
spherical stellar wind or a spherical homogeneous UC H\,{\sevensize II}
region with an excess of dust emission at 2.7 mm. If interpreted as an
UC \hii region, the size of the \hii region is only 2.5 mpc and is
excited by a B1 ZAMS star \citep{hofner1996}. \citet{franco2000} quote a
slightly flatter spectral energy distribution between 8.4 and 110 GHz,
with a spectral index of $0.95 \pm 0.06$ which they attribute to a
density gradient proportional to $r^{-2.5}$. Using the 1.3 cm data of
\citet{testi2000} for G9.62+0.20E, \citet{franco2000} derives an
electron density of about $6 \times 10^5~\mathrm{cm^{-3}}$ which could
rise to above $10^6~\mathrm{cm^{-3}}$ in a uniform core region. Taken
together with the size of the \hii region, G9.62+0.20E qualifies as a
hypercompact \hii region, implying a very early phase of massive star
formation.

Component E is also well defined in a number of sulphur and nitrogen
bearing molecules as well as in some organic molecules
\citep{su2005}. Near-infrared imaging of the star forming region shows
emission at components B, C and E \citep{testi1998}. \citet{persi2003}
subsequently found that the NIR source near component E has the colours
of a foreground star. \citet{debuizer2005} find weak emission at 11.7
$\mu$m towards the radio component E.

In this paper we present further monitoring data on G9.62+0.20E for
methanol masers at 6.7, 12.2, and 107 GHz. We present various analyses
of the data and consider a possible explanation for the variability
of the masers. 

\section[]{Observations and data reduction}
\label{sec-obs}
{\it HartRAO observations:} Observations at 6.7 GHz and 12.2 GHz were
made with the 26-m Hartebeesthoek telescope.  Both receivers provide
dual circular polarization.  The system temperatures at zenith are
typically 60 K at 6.7 GHz and 100 K at 12.2 GHz.  The spectrometer
provides 1024 channels per polarization, and the bandwidths used for
spectroscopy of 1 MHz at 6.7 GHz and 2 MHz at 12 GHz provide spectral
resolutions of 0.044 and 0.048 km s$^{-1}$ respectively.

For spectroscopic observing, the telescope pointing error was determined
through five short integrations at the cardinal half power points of the
beam and on source.  The on-source spectra were scaled up by determining
the amplitude correction from the pointing error using a Gaussian model
for the main beam.

Calibration was based on monitoring of 3C123, 3C218 and Virgo A (which
is bright but partly resolved), using the flux scale of \citet{ott1994}.
The antenna temperature from these continuum calibrators was measured by
drift scans.  Pointing errors in the north-south direction were measured
via drift scans at the beam half power points, and the on-source
amplitude was corrected using the Gaussian beam model.  The bright
methanol maser source G351.42+0.64, which previous monitoring has shown
to exhibit only small variations in the strongest features
\citep{goedhart2004}, was observed in the same way as G9.62+0.20 to
provide a consistency check on the spectroscopy.  These sources were
observed frequently, usually on the same days as G9.62+0.20.

{\it ARO 12m observations:} The observations with the ARO 12m
telescope\footnote{The Kitt Peak 12 Meter telescope is operated by the
Arizona Radio Observatory (ARO), Steward Observatory, University of
Arizona. } were done from December 3, 2003 to June 25, 2004 to observe
cycle 8 and from January 14, 2005 to April 22, 2005 to observe cycle
9. The peak for cycle 8 was expected in early July 2004. Poor
atmospheric conditions forced any further observations to be abandoned
after June 2004. The onset of flare 9 was expected to occur around
February 7, 2005 with a peak between March 7 - 13 and to decay to its
low state by middle April 2005. The receivers used were dual-channel SIS
mixers operated in single-sideband mode. The backends were filter banks
with 100kHz and 250 KHz resolution. The observing frequency was
107.01385 GHz.  Each observing run lasted about three hours, during
which regular positional checking on planets were done.

\begin{figure}
\resizebox{\hsize}{!}{\includegraphics[width=84mm,clip,angle=-90]{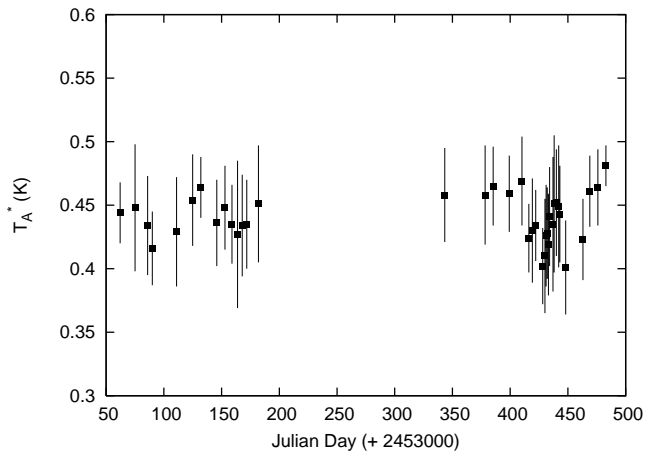}}
\caption{Time series for the continuum emission of DR-21 at 107 GHz}
\label{fig:dr21_ts}
\end{figure}

DR-21 was used as calibrator as well as a reference source and was
observed in spectral line mode.  Data reduction was done with the CLASS
package. Fig.\ref{fig:dr21_ts} shows the time series for DR-21.
The results are for the 250 KHz resolution and each point is the average
over all channels. We estimated the flux density
of DR-21 at 107 GHz using the continuum spectrum given by
\citet{righini1976}, which gives a flux density of 16.7 Jy. For each
observing run an appropriate conversion factor was calculated and
applied to the corrected antenna temperature of G9.62+0.20E.

\section{Results and analysis}

\begin{figure}
\resizebox{\hsize}{!}{\includegraphics[width=84mm,clip,angle=0]{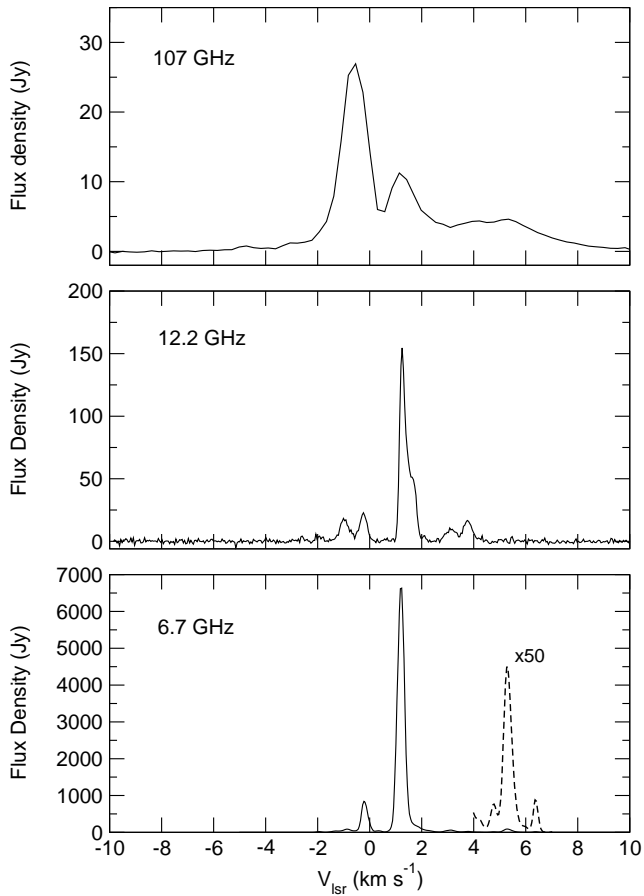}}
\caption{Representative spectra for the masers at 6.7, 12.2 GHz. The 107
  GHz spectrum is an average taken over all observations. The features
  at 5.3 and 6.4 $\mathrm{km~s^{-1}}$ in the 6.7 GHz spectrum are
  associated with G9.62+0.20D.}
\label{fig:avgspec}
\end{figure}

The average or representative spectra for the three maser transitions
are shown in Fig. \ref{fig:avgspec}. It is seen that, whereas for the
6.7 and 12.2 GHz spectra the strongest feature is located at about 1.3
$\mathrm{km~s^{-1}}$, the corresponding feature in the 107 GHz spectrum
is the weaker maser feature. The 107 GHz spectrum also shows a broad
thermal component. G9.62+0.20E has in the past been observed at 107 GHz
by \citet{valtts1995, valtts1999} and \citet{caswell2000}.  The average
107 GHz spectrum in Fig. \ref{fig:avgspec} resembles the result of
\citet{caswell2000}. The observations of \citet{valtts1995, valtts1999}
show only one maser feature at -0.57 $\mathrm{km~s^{-1}}$ and also no
broad thermal component.

\begin{figure*}
\centering
\includegraphics[width=200mm,clip,angle=270]{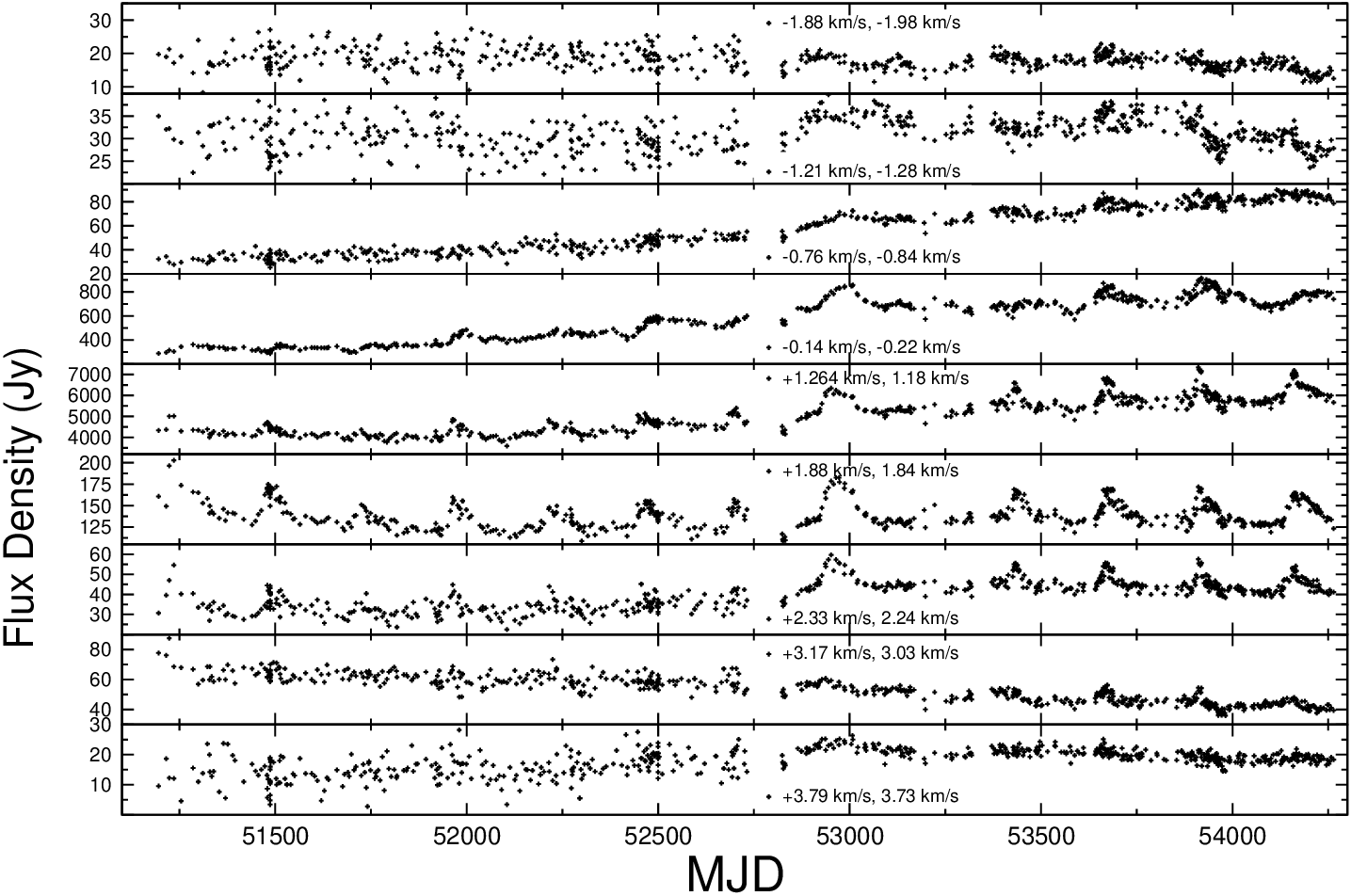}
%\vbox to220mm{\vfil 6.7 GHz time series to go here
\caption{6.7 GHz time series}
\label{fig:67ts}
%\vfil}
\end{figure*}

\begin{figure*}
\centering
\includegraphics[width=200mm,clip,angle=270]{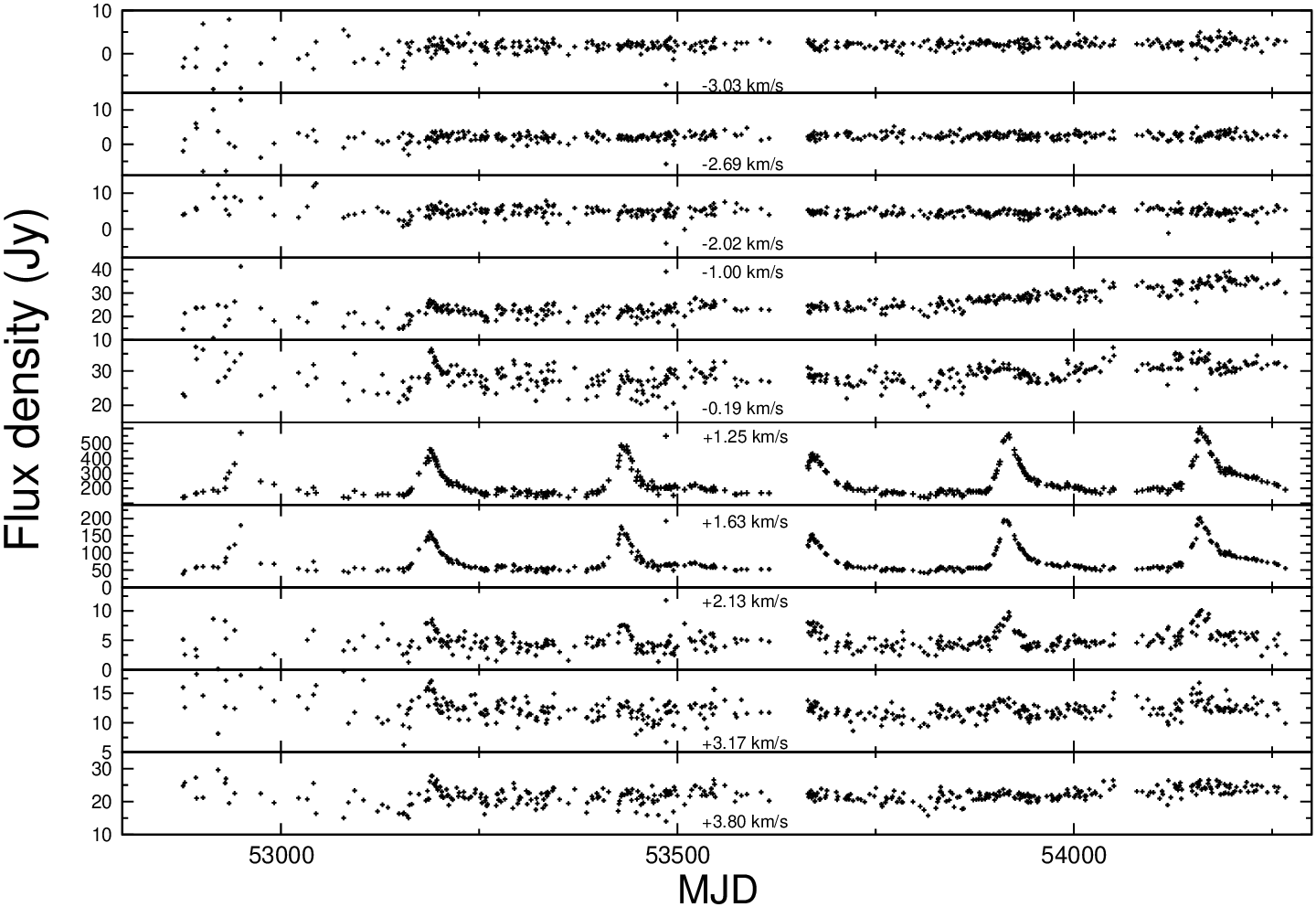}
%\vbox to220mm{\vfil 12.2 GHz time series to go here
\caption{12.2 GHz time series}
\label{fig:122ts}
%\vfil}
\end{figure*}

The full time series  for all the identifiable features in the 6.7 and
12.2 GHz spectra are shown in Figs. \ref{fig:67ts} and
\ref{fig:122ts}. For the 6.7 GHz maser, eleven different features could be
identified in the single dish spectrum and ten for the 12.2 GHz
maser. The time series covers about 2670 days and 13 flares. Inspection
of Fig. \ref{fig:67ts} shows that while there are some similarities in
the time series of some of the 6.7 GHz maser features, there also are
some significant differences. 

The strongest varying 12.2 GHz masers are significantly more
variable than the strongest varying 6.7 GHz masers. To quantify the
variability we calculated the relative amplitude, defined as
\begin{equation}
R = \frac{S_{max} - S_{min}}{S_{min}} = \frac{S_{max}}{S_{min}} - 1
\label{eq:r}
\end{equation} 
for a number of features in the three transitions.  The results are
given in Table \ref{tab:amplitudes}. The errors indicate the spread in
the relative amplitudes for the last six flares in the time series. No
error could be given for the single 107 GHz flare that has been
monitored. The 12.2 GHz
masers are indeed significantly more variable than the 6.7 and 107 GHz
masers while the relative amplitudes of the 6.7 and 107 GHz masers are
approximately the same.

\begin{table*}
\begin{minipage}{120mm}
\caption{Relative amplitudes (eq. \ref{eq:r}) for selected masers at 6.7, 12.2 and 107 GHz}
\label{tab:amplitudes}
\begin{tabular}{cccccc}
\hline
\multicolumn{2}{c}{6.7 GHz} & \multicolumn{2}{c}{12.2 GHz} & \multicolumn{2}{c}{107 GHz} \\
\hline
Velocity & Relative & Velocity & Relative & Velocity & Relative \\
$\mathrm{km\, s^{-1}}$ & amplitude & $\mathrm{km\, s^{-1}}$ & amplitude& $\mathrm{km\,
  s^{-1}}$ & amplitude\\
1.18 & 0.26 $\pm$ 0.04 & 1.25 & 2.0 $\pm$ 0.4 & 1.14 & 0.31 \\
1.84 & 0.32 $\pm$ 0.05 & 1.63 & 2.4 $\pm$ 0.3 &  & \\
2.24 & 0.33 $\pm$ 0.04 & & & & \\
\hline
\end{tabular} 
\label{table:r}
\end{minipage}
\end{table*}

Figures \ref{fig:67ts} and \ref{fig:122ts} also show that the regular
flaring does not occur in  all of the 6.7 and 12.2 GHz maser
features.  For the 12.2 GHz masers it is seen that the flaring behaviour
is limited to three features namely $+1.25 ~\mathrm{km\:s^{-1}}$, $+1.63
~\mathrm{km\:s^{-1}}$ and $+2.13~ \mathrm{km\:s^{-1}}$. The strongest
flaring occurs at +1.25 and +1.63 $\mathrm{km\:s^{-1}}$. These are also
the strongest two features in the maser spectrum. The absence of flaring
behaviour outside the above velocity range is quite obvious. Because of
the smaller variability of the 6.7 GHz masers the question of which
features are flaring is not as simple as for the 12.2 GHz masers. Simple
visual inspection suggests that flaring occurs for the features at -0.2,
+1.2, +1.8, and +2.2 $\mathrm{km\:s^{-1}}$.

It is noteworthy that the most variable and brightest 12.2 GHz masers, at
+1.63 and +1.25 $\mathrm{km\:s^{-1}}$, lie close to each other at the
northern end of a north-south chain of maser features as seen in the
high resolution maps of \citet{goedhart2005b}. As already pointed out,
these are also the two brightest masers in the 12.2 GHz spectrum. The
maser features that show no periodic flaring lie significantly offset to
the south from the above two features. 

We also checked for periodic variations in the light curves using the
Lomb-Scargle periodogram \citep{scargle1982} and Davies' L-statistic
\citep{davies1990}. For the 12.2 GHz masers both methods confirm the
existence of a periodic signal with period 244 days in the light curves
of the $+1.25~\mathrm{km\:s^{-1}}$, $+1.63 \mathrm{km\:s^{-1}}$ and
$+2.13~\mathrm{km\:s^{-1}}$ features at a very high level of
significance. For the 6.7 GHz masers the periodograms were significantly
more complicated than that of the 12.2 GHz masers. A full analysis and
discussion of the periodograms falls outside the scope of the present
paper and will not be discussed here. Within the context of the present
paper it is necessary to note that for the 12.2 GHz masers,
periodic (regular) flaring is observed only in three features. For the
6.7 GHz masers, flaring behaviour is strongest also for only a subset of
the maser features, some of which coincide in velocity with the 12.2 GHz
masers that show flaring behaviour.

\begin{figure}
\centering
\includegraphics[width=84mm,clip,angle=0]{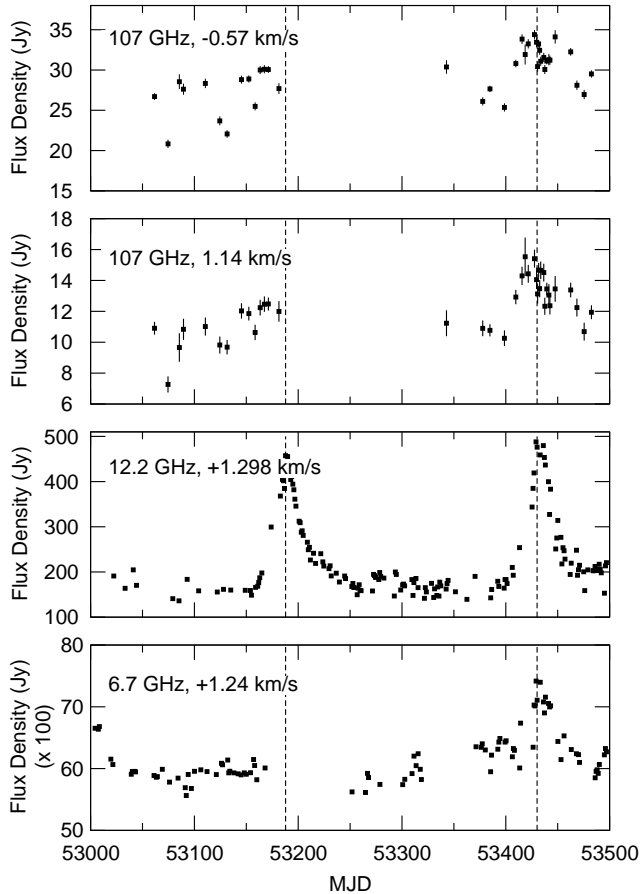}
\caption{Time series for masers at 6.7, 12.2 and 107 GHz. For the 6.7
  and 12.2 GHz masers only one velocity component has been
  considered. The solid dashed vertical lines indicate the position of
  the peak of the 12.2 GHz maser.}
\label{fig:ts}
\end{figure}

In Fig. \ref{fig:ts} we show the time series  for cycles 9 and 10 for
which the 107 GHz masers have also been monitored. The vertical dashed
lines give the position of the peak of the flare at 12.2 GHz for the
1.25 $\mathrm{km\:s^{-1}}$ feature. For cycle 9 the scatter in the 107
GHz data is quite large and, although there might be a slight hint of a
rise in the flux density towards the expected time when the flare was at
its peak, no clear sign of a flare can be identified. The reason for the
large scatter in the data is most probably the poor atmospheric
conditions that prevailed at that time and which led to the termination
of the monitoring for cycle 9.

\begin{figure}
\includegraphics[width=85mm,clip,angle=0]{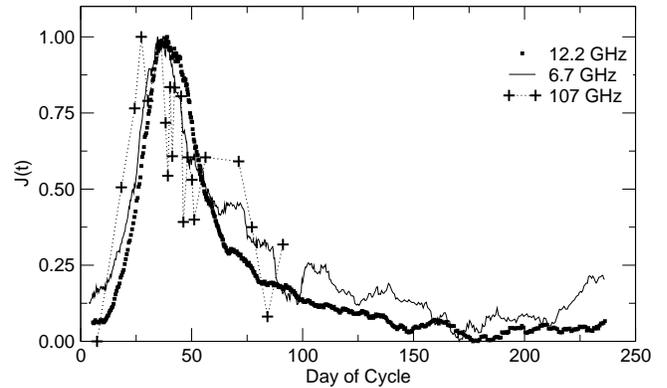}
\caption{Comparison of the average flare profile for the 1.24
  $\mathrm{km~s^{-1}}$ feature for the 6.7, 12.2, and 107 GHz
  masers. For the 6.7 and 12.2 GHz masers the flare profile was obtained
  from a 30 point running average.}
\label{fig:flareprofile}
\end{figure}

Closer inspection of the time series in Figs. \ref{fig:67ts},
\ref{fig:122ts}, and \ref{fig:ts} suggests that the flare profiles for
the 6.7, 12.2 and 107 GHz masers might be different. To investigate to
what extent the flare profiles are similar we first considered the
average flare profiles by calculating the normalized profile
\begin{equation}
J(t) = \frac{S(t) - S_{min}}{S_{max} - S_{min}}
\label{eq:jt}
\end{equation} 
for the +1.25 $\mathrm{km\:s^{-1}}$ feature of the 12.2 GHz spectrum and
for the corresponding feature at 6.7 GHz. For both these features a 30
point running average was used after detrending the time series and
folding of the data modulo the period of 244 days. The 107 GHz data for
the one flare were scaled accordingly. The results are shown in
Fig. \ref{fig:flareprofile}.  In spite of the fact that the average
profiles do not coincide exactly in time, it is remarkable to note that
the profiles seem to be very similar for the three maser transitions.

\begin{figure}
\centering
\includegraphics[width=85mm,clip]{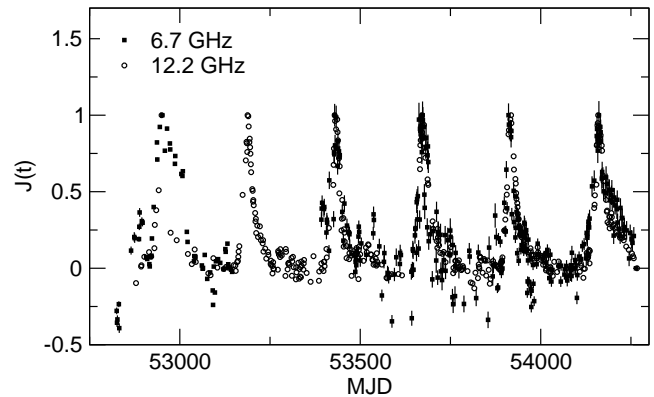}
\caption{Comparison of the individual normalized flare profiles for the
  1.24 $\mathrm{km~s^{-1}}$ feature for the 6.7 and 12.2 GHz masers for
  MJD $>$ 52750}
\label{fig:jt}
\end{figure}

The same analysis as above was also carried out on individual 6.7 and
12.2 GHz flares to examine to what extent individual flares have the
same profile. The result is shown in Fig. \ref{fig:jt}. For clarity only
the flares that occured after MJD 52750 are shown and only for the
features at 1.25 $\mathrm{km\:s^{-1}}$. In spite of the large errors on
the 6.7 GHz data, it is seen that the flare profiles are basically the
same, also for individual flares. The similarity of the individual
  flare profiles suggests that the characteristics of the mechanism
  underlying the flaring remain the same from flare to flare and that it
  affects the 6.7 and 12.2 GHz masers in the same way.

We finally note that within the velocity resolution of the observations,
no velocity shifts of the different feature in the maser spectrum could
be detected during the flares.

\section{Discussion}
\citet{goedhart2005a} briefly considered a number of possible mechanisms
that might underlie the flaring behaviour of the masers in
G9.62+0.20E. These included: disturbances like shock waves or clumps of
matter passing through the masing region; variations in the flux of seed
photons or of pump photons due to stellar pulsations; periodic
outbursts, or the effects of a binary system. While it is not possible
to come to a final conclusion about the origin of the periodicity of the
masers with the present data, it is possible to at least exclude certain
mechanisms and point to possible explanations for the periodicity.

It has already been noted that the return of the masers to basically the
same quiescent level between flares is a strong indication that the
masing regions remain unaffected by whatever mechanism underlies the
flaring. Mechanisms such as shock waves or clumps of matter passing
through the masing region are therefore excluded. This implies that
whatever the underlying mechanism for the periodicity, the coupling with
the masing regions must be radiative, ie. either through the seed
photons or through heating and cooling of the dust that is responsible
for the pumping radiation field.

At present the information available on the masers are the single dish
light curves (as presented above) and the high resolution
interferometric mapping of \citet{goedhart2005b}.  We consider the light
curve (flare profile) only since in general the light curves of periodic
sources contain significant information about the physical processes
that are responsible for the observed variability. We also present an
analysis that strongly suggests that the decay part of the light curve
might be due to the background \hii region decaying from a higher to a
lower state of ionization. It is necessary to point out here that since
the relative positions of the masers with respect to the \hii region is
not known to milli-arcsecond accuracy, a more complete understanding of
all aspects of the variability of the masers in G9.62+0.20E cannot be
given here.

For the present discussion we will focus on the strongest maser features
and take the average flare profile of the 12.2 GHz masers
(Fig. \ref{fig:flareprofile}) as representative of all flares. The flare
profile can be described by a rather steep rise followed by a decaying
part lasting about 100 days to eventually reach a minimum (or possibly
quiescent) state after which it appears to slowly increase before the
next flare starts again. Since, as we have already argued that effects
that physically affect the masing region can be ruled out, we are
basically left with two possibilities to understand the light curve:
stellar pulsation and a binary system.

A comparison of the maser light curves as described above with the light
curves of all classes of pulsating stars clearly shows that the maser
light curve has a completely different behaviour from that of any class
of pulsating stars. In fact, considering the physics of stellar
pulsation it is hard to see how to produce a light curve for a pulsating
star such that it resembles that of the masers in G9.62+0.20E. For the
masers the light curve is strongly suggestive of a flaring state that is
superposed on top of a base level emission. Pulsating stars are not in an
equilibrium state and therefore do not have such a quiescent state on
top of which pulsations or flares are superposed.  Explaining the
observed maser light curves as being due to pulsation of the central
star would therefore appear to be very difficult.

\begin{figure}
\centering
\includegraphics[width=80mm,clip,angle=0]{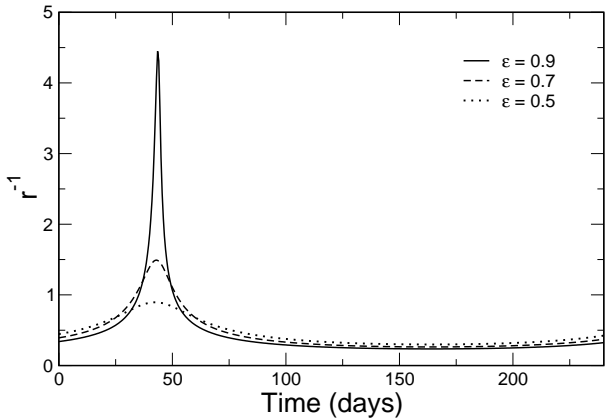}
\caption{Time dependence of $r^{-1}$ for different eccentricities. }
\label{fig:profile}
\end{figure}

The only alternative is that of a binary system. The light curve does
not suggest that we are dealing with an eclipsing effect for the masers.
A class of binary systems which potentially can meet the requirement of
providing radiative coupling between the source of variability and the
masing regions is the colliding-wind binaries.  In fact,
\citet{zhekov1994} argued that, given the high binary frequency found in
young stars and the observed mass loss rates and wind velocities,
supersonically colliding-wind systems, should also be found amongst
pre-main sequence binary systems.  Although it is not possible with the
available data to come to a conclusion here whether or not G9.62+0.20E
indeed harbours a colliding-wind binary, it is worth exploring this
possibility somewhat more by considering some properties of
colliding-wind binaries.  

The main ingredient of a colliding-wind binary is the two oppositely
facing shocks separated by a contact discontinuity. The postshock
temperature is given by $T_{sh} = 3mv_w^2/16k$ \citep{stevens1992,
  zhekov1994}, where $m$ is the mean mass of the particles that
constitute the wind, and $v_w$ the wind speed. Given that the central
star for G9.62+0.20E is estimated to be of spectral type B1 and using
the results of \citet{bernabeu1989}, a wind speed of
$800~\mathrm{km~s^{-1}}$ does not seem to be an unrealistic
assumption. Using this value for $v_w$ and assuming for simplicity a
pure hydrogen wind, a postshock temperature of $\sim 1.5 \times 10^7$ K
is found. \citet{pittard2005} presented the emissitivity for hot thermal
plasmas of temperatures $10^6$, $10^7$, and $10^8$ K, from which an
extrapolation to energies below 100 eV suggests that the post-shock gas
will emit photons from the visible up to X-ray energies. Using the
CHIANTI code \citep{dere1997, landi2006}, we confirmed that this is
indeed the case. This range of photon energies obviously includes
photons that can heat the inner edge of the circumstellar dust as well
as ionizing photons that can cause additional ionization in the \hii
region, respectively affecting the pumping radiation field and/or the
background source of seed photons.  Due to the lack of a numerical model
it was not possible to calculate expected absolute fluxes of ionizing
photons produced at the shocks as well as of photons that can heat the
dust. A quantitative evaluation of  the effects these photons might
have on the \hii region and the dust was therefor not possible.

A second property of colliding-wind systems relevant to this discussion,
is that the total luminosity at the shock scales like $r^{-1}$, where
$r$ is the distance between the two stars \citep{stevens1992,
  zhekov1994}. Obviously this scaling property implies a modulating
effect of the total luminosity at the shock during the orbital motion
for eccentric orbits.  To investigate the time dependence of $r^{-1}$,
we used Kepler's third law to calculate the semi-major axis for the
binary system. For this we used a mass of 17 $\msun$ for the ionizing
star (spectral type B1) and arbitrarily adopted a mass of 8 $\msun$ for
the secondary star. Since in Kepler's third law the semi-major axis
depends on the inverse of the third root of the total mass, the final
answer is not very sensitive to the mass of the secondary. Using the
period of 244 days it then follows that the semi-major axis of the orbit
is 2.23 AU. For an elliptic orbit the radial distance between the two
stars is given by $r = a(1-e^2)/(1 + e\cos\theta)$ with $a$ the
semi-major axis and $e$ the eccentricity. The resulting orbital
modulation, for eccentricities of 0.5, 0.7, and 0.9, of the luminosity
is shown in Fig. \ref{fig:profile} with $r$ in AU. The position of
periastron passage has been set at 43.5 days in all three cases. 

Combining the above two basic properties of colliding-wind binary
systems, it is clear that such systems can provide a periodic source of
photons that can heat the circumstellar dust and thereby possibly affect
the pumping radiation field for the masers, as
well as ionizing photons that can cause additional ionization in the
\hii region. Although the exact projection of the masers against the
\hii region in G9.62+0.20E is not known, the work of
\citet{phillips1998} suggests that the \hii region might indeed be the
source of seed photons for the masers.

Comparison of Fig. \ref{fig:profile} with the maser flare profile,
however, shows that, even in the case of an eccentricity of 0.5, the
observed maser flare profile has a decay time that is significantly
longer than what is expected simply from the radiation pulse produced at
periastron passage.  Within the framework of the colliding-wind binary
scenario, the observed decay must then be due to either the cooling of
dust or the recombining of the \hii region (or parts thereof) from a
higher degree of ionization to its pre-flare state.

Using the fact that the dust cooling time is proportional to $T^{-6}$
\citep{kruegel2003} for the optical thin case and proportional to
$T^{-4}$ in the optically thick case, and that in the optical thin case
the typical cooling times for small grains with a temperature of 60 K is
about 10 seconds \citep{kruegel2003}, an upper limit of about 10 hours
for the cooling time is found for the optically thick case. Assuming now
a grain temperature of 100K and that the cooling time under optically thin
conditions is also 10 seconds, which, due to the $T^{-6}$ dependence
should actually be less, a cooling time of 1.2 days is found for the
optically thick case. Thus, even in the optically thick case the dust
cooling time is only a small fraction of the decay time of the maser
flare, suggesting that the decay of the maser flare is most probably not
due to the cooling of dust.

To investigate the possibility that the decaying part of the maser flare
is due to the recombination of the \hii region or parts thereof, we note
that the recombination time scale of a hydrogen plasma is given by
$(\alpha n_e)^{-1}$ \citep{osterbrock1989}, with $\alpha$ the
recombination coefficient and $n_e$ the electron density. Using $\alpha
= 2.95 \times 10^{-13}~\mathrm{cm^3s^{-1}}$, characteristic decay times
between 40 and 400 days are found for densities between $n_e =
10^6~\mathrm{cm^{-3}}$ and $n_e = 10^5 ~\mathrm{cm^{-3}}$. This is
significantly longer than the dust cooling time and seems to be able to
account for the observed decay time of about 100 days.

\begin{figure}
\centering
\includegraphics[width=80mm,clip,angle=0]{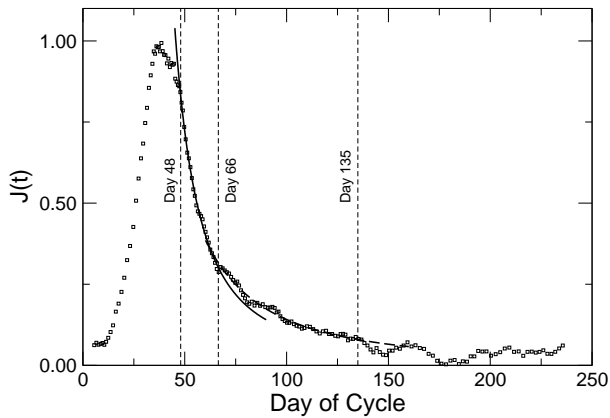}
\caption{Normalized average flare profile for the 12.2 GHz masers with
  the solid and dashed lines the fits from eq. \ref{eq:decay}. The vertical
  dashed lines indicate the time intervals which was used for the
  fitting. }
\label{fig:hiidecay}
\end{figure}

In view of these numbers, the question arises whether the decay of
  the maser flare can indeed be ascribed to a change in the seed photon
  flux from a recombining thermal plasma. In Appendix A we derive an
  expression for the time dependence of the electron density in a volume
  of ionized hydrogen going from a higher level of ionization to a lower
  equilibrium ionization state, due to the recombination of the plasma.
  It is shown that in the optically thin case, the intensity of the
  associated free-free emission decreases with time according to
\begin{equation}
I_\nu(t)  \propto  \left[\frac{1 + u_0\tanh(\alpha n_{e,\star}t)}{u_0 + \tanh(\alpha
    n_{e,\star}t)}\right]^{-2} \label{eq:decay}
\end{equation}
with $u_0 = n_{e,0}/n_{e,\star} > 1$.  $n_{e,0}$ is the electron density
from where the decay starts at time $t = 0$, and $n_{e,\star}$ the
equilibrium electron density determined by ionization balance for the
ionizing stellar radiation. Fitting eq. \ref{eq:decay} to the decay part
of the flare will give an indication to what extent the decay of the
maser flare can be interpreted as being due to the recombination of a
hydrogen plasma. An initial non-linear regression analysis using the
above time dependence suggested that the decay of the maser flare
actually consists of two components that have to be fitted
separately. The first component is from days 48 to 66, and the second
from days 67 to 135.

Following the discussion in Appendix A, we thus fitted the two
components with eq. \ref{eq:decay} by systematically varying
$n_{e,\star}$. This allowed us to estimate $u_0$ and therefore $n_{e,0}
= u_0n_{e,\star}$, ie. the electron density at the time when the
ionization pulse was switched off or when its effect was not significant
anymore. For the interval between days 48 and 66 it was found that
$n_{e,0}$ varies from $1.57 \times 10^6~\mathrm{cm^{-3}}$ to $2.00
\times 10^6~\mathrm{cm^{-3}}$ for $n_{e,\star}$ ranging from
$1.0\times10^5~\mathrm{cm^{-3}}$ to $7.0 \times 10^5~\mathrm{cm^{-3}}$.
Similarly, for the interval between days 67 and 135, $n_{e,0}$ varies
between $6.0 \times 10^5~\mathrm{cm^{-3}}$ and $7.3 \times
10^5~\mathrm{cm^{-3}}$ for $n_{e,\star}$ ranging from $1.0 \times
10^4~\mathrm{cm^{-3}}$ to $2.0 \times 10^5~\mathrm{cm^{-3}}$. For each
interval the lower value for the range of $n_{e,\star}$ was chosen such
that $n_{e,0}$ is also the solution of eq. \ref{special}, ie. the
solution for the limiting case when $n_{e,\star} \ll n_{e,0}$. The upper
value for the range of $n_{e,\star}$ was chosen to be about one third or
half of the value of $n_{e,0}$ obtained for the lower value of the range
of $n_{e,\star}$.

In Fig. \ref{fig:hiidecay} we show the fit of eq. \ref{eq:decay} to
  each of the two time intervals of the decay of the maser flare for
  specific values of $n_{e,\star}$ from the abovementioned ranges of
  values. For days 48 to 66 we used $n_{e,\star} = 3 \times
10^5~\mathrm{cm^{-3}}$ and for days 67 to 135 we used $n_{e,\star} = 1
\times 10^5~\mathrm{cm^{-3}}$. These two values correspond to $n_{e,0} =
1.66 \times 10^6~\mathrm{cm^{-3}}$ and $n_{e,0} = 6.36 \times
10^5~\mathrm{cm^{-3}}$, respectively. The quality of the fits is
  quite remarkable, especially given that a priori there is no
  information suggesting that the decay of the maser flare might
  follow the recombination of a thermal hydrogen plasma. The values for
  $n_{e,0}$ as estimated from the non-linear regression are also what is
  expected for hypercompact \hii regions. In fact, the agreement with the
  densities quoted by \citet{franco2000} for G9.62+0.20E is also
  remarkable. At the least these results are strongly suggestive that
the decay of the maser flare might be due to the recombination of an
\hii region, or parts thereof, from a higher to a lower ionization
state. The entire maser flare might therefore be due to changes in the
background source.

We also note that within the framework of the hypothesis that the maser
flaring is due to changes in the background \hii region, day 48 gives an
upper limit of the time when the radiative pulse causing the increase in
the ionization level, has ``switched'' off and the \hii region is left
to decay to its equilibrium level. Referring to
Fig. \ref{fig:flareprofile} the start of the flare is at about day 10,
implying that the full width of the radiative pulse is at most about 38
days. This suggests a rather sharply peaked pulse which is what is
qualitatively expected from the colliding-wind binary scenario discussed
above.

To what extent the colliding-wind binary scenario is the only one that
can explain the maser flaring within the broader framework of binary
systems, is not clear yet. Our discussion above does not include the
presence of a possible accretion disk or even two accretion disks, as
well as outflows. Obviously such scenarios are much more complex, but
should nevertheless be able to explain the periodic flaring as well as
the shape of the flare profiles. Given the short cooling time of the
dust, it follows, if the entire flare profile is due to changes in the
pumping radiation field, that the primary driving source of the
variable infrared radiation field should basically have the same time
dependence as the maser flare profile. Furthermore, the physical process
should be rather stable from orbit to orbit to explain the similarity of
the flare profiles in the time series. Obviously, monitoring of
G9.62+0.20E in the mid-infrared might help resolve this problem.

\section{Summary and Conclusion}

We presented the results of the monitoring of methanol masers in
G9.62+0.20 at 6.7, 12.2, and 107 GHz. Like the 6.7 and 12.2 GHz masers,
the 107 GHz masers also show flaring behaviour, with a relative amplitude
that is the same as that of the 6.7 GHz masers. It was also found that
the flare profile is the same for the 6.7, 12.2 and 107 GHz masers, and
that for the 6.7 and 12.2 GHz masers the profiles of individual flares
also are the same.  We have also shown that in the low phase of the
masers, i.e. between two flaring events, the ratios of maser flux
densities return to the same levels. From this behaviour we conclude
that the physical conditions in the masing region must be relatively
stable and that the source for the flaring behaviour lies outside the
masing region. 

Comparison of the maser light curves with that of pulsating stars lead
us to conclude that stellar pulsation is not the underlying cause for
the observed maser flaring. This leaves a binary system as the only
other option.  It was argued that, at least qualitatively, a
colliding-wind binary can provide the mechanism for a periodic
background source and/or a periodic pumping radiation field.  Although
there might be heating of the dust due to radiation from the shocked
regions, and thus the continuum of G9.62+0.20E might then show
variability in the IR, it seems difficult to explain the decay time of
about 100 days as being due to the cooling of dust.

It was shown that the characteristic recombination time of an \hii
region with densities in the range $10^5~\mathrm{cm^{-3}}$ to
$10^6~\mathrm{cm^{-3}}$ can explain the observed decay time of the maser
flare. We also showed that during the intervals 48 to 66 days and 67 to
135 days, the decay of the maser flare is what can be expected for the
recombination of thermal plasmas with densities of approximately $1.6
\times 10^6~\mathrm{cm^{-3}}$ and $6.0 \times 10^5~\mathrm{cm^{-3}}$,
respectively. These values are in very good agreement with densities
derived independently from radio continuum measurement.

A number of questions, however, still need to be answered. Can we find
observational evidence for a colliding-wind binary?  Is there a periodic
infrared, radio continuum, or X-ray signal associated with G9.62+0.20E?
Is the ionizing photon flux produced in the hot post-shock gas large
enough that it still can have an observable effect on \hii region in
terms of changes in the free-free emission? Are there other mechanisms
associated with young binary systems that can also account for the maser
flaring? It is also necessary to determine the exact position of the
masers relative to the \hii region.  Obviously, significantly more work
still needs to be done before a final answer on the periodic flaring in
G9.62+0.20E can be given.

\section*{Acknowledgements}
DJvdW was supported by the National Research Foundation under Grant
number 2053475. We would like to thank Moshe Elitzur and Andrei
Ostrovskii for valuable discussions in the early part of this
project. DJvdW also would like to acknowledge discussions with Karl
Menten and Endrik Kr\"ugel.

\bibliographystyle{mn2e} 
\bibliography{mnrefs} 

\appendix

\section[]{The decay of the electron density in an \hii region} We
consider a volume element of a pure hydrogen \hii region which is not
fully ionized. The ionization rate due to the stellar radiation is given
by $\Gamma_\star$. The total hydrogen density is $n_H$ while the
electron density due to ionization by the stellar radiation field is
given by $n_{e,\star}$, the neutral hydrogen density by $n_{H^0}$ and,
$n_H = n_{e,\star} + n_{H^0}$. We now add an additional time-dependent
source of ionizing radiation, described by a time-dependent ionization
rate $\Gamma_p(t)$. The rate equation for the electron density is then
given by 
\begin{equation}
\frac{dn_e}{dt} = -\alpha n_e^2 + (\Gamma_{\star} + \Gamma_p(t))n_H
\end{equation}
where $\alpha$ is the recombination coefficient.  The first term on the right
gives the decrease of the electron density due to recombinations and the
second term the production of electrons due to photoionizations.

At time $t = 0$ the additional source of ionization is switched off, and
$n_e = n_{e,0} > n_{e,\star}$.  The equation that then governs the recombination of
the plasma is
\begin{equation}
\frac{dn_e}{dt} = -\alpha n_e^2 + \Gamma_{\star}n_H 
\end{equation}
Recombination and ionization balance for the stellar radiation 
implies $\Gamma_{\star}n_H = \alpha n^2_{e,\star}$. The rate equation
then becomes
\begin{equation}
\frac{dn_e}{dt} = -\alpha n_e^2 + \alpha n^2_{e,\star}
\label{ricatti} 
\end{equation}
This is a Riccati equation for which analytical solutions can be
found. For eq. \ref{ricatti} the solution is given by
\begin{equation}
n_e(t) = n_{e,\star}\left[\frac{u_0 + \tanh(\alpha n_{e,\star}t)}{1 + u_0\tanh(\alpha n_{e,\star}t)}\right]
\label{general}
\end{equation}
where $u_0 = n_{e,0}/n_{e,\star} > 1$. It is straight forward to check
that at $t = 0$, $n_e = n_{e,0}$ and that for large $t$, $n_e(t)
\rightarrow n_{e,\star}$. 

We note that a special case of eq. \ref{ricatti} is when $n_{e,\star}=
0$. The second term in eq. \ref{ricatti} can then be dropped. In this
case the solution is easily found to be
\begin{equation}
n_e(t) = \frac{n_{e,0}}{1+\alpha n_{e,0} t}
\label{special}
\end{equation}
This solution can also be found from eq. \ref{general} when $n_{e,\star}
\ll n_{e,0}$.

Since we do not measure the electron density directly but rather the
intensity of free-free emission from the plasma, we need to calculate
$I_\nu(t)$. We can consider the two extreme cases of the \hii region
being optically thick or optically thin. In the optically thick case we
have $I_\nu = B_\nu(T)$ which is independent of the electron density and,
if $T$ is constant, it follows that $ dI_\nu/dt = 0 $.

In the optically thin case we have
\begin{equation}
I_\nu(t) = \tau_\nu(t) B_\nu(T) \propto n_e^2(t) B_\nu(T)
\end{equation}
since $\tau \propto n_e^2$. For the general case (eq. \ref{general}) it
follows therefore that
\begin{equation}
I_{\nu}(t) \propto n^2_{e,\star}\left[\frac{1 + u_0\tanh(\alpha n_{e,\star}t)}{u_0 + \tanh(\alpha
    n_{e,\star}t)}\right]^{-2} \label{eq:decay2}
\end{equation}
while for the special case (eq. \ref{special}) 
\begin{equation}
I_\nu(t) \propto n_{e,0}^2(1 + \alpha n_{e,0} t)^{-2} 
\end{equation}
In terms of logarithms we have
\begin{equation}
\log I_\nu(t) \propto -2\log\left[\frac{1 + u_0\tanh(\alpha n_{e,\star}t)}{u_0 + \tanh(\alpha
    n_{e,\star}t)}\right]
\label{general2}
\end{equation}
and
\begin{equation}
\log I_\nu(t) \propto -2\log(1 + \alpha n_{e,0} t)
\label{special2}
\end{equation}
for the general and special cases respectively.

When doing a regression analysis using eq. \ref{eq:decay2}, it is
  obviously better to resort to the logarithmic representation as in
  eq. \ref{general2}. However, when written as an equality,
  eq. \ref{general2} has three unknown parameters: a constant, $u_0$ and
  $n_{e,\star}$. Since a linear regression can only give the values of
  two unknowns, it is necessary to specify a value for one of the three
  unknowns. In the present paper we specify the value of $n_{e,\star}$
  since it is known what realistic values $n_{e,\star}$ can assume. This
  allows us then to estimate $u_0$ and therefor also $n_{e,0}$.

\bsp

\label{lastpage}

\end{document}